\documentclass{kapproc} 

\usepackage{procps} 

\usepackage[dvips]{graphicx}

\upperandlowercase

\normallatexbib 

\begin{document}

\articletitle{Optical properties of 
(P\MakeLowercase{r},C\MakeLowercase{e})$_2$C\MakeLowercase{u}O$_4$}

\author{A. Zimmers,$^1$ N. Bontemps,$^1$ R.P.S.M. Lobo,$^1$ 
C.P. Hill,$^2$ M.C. Barr,$^2$ \\
R.L. Greene,$^2$ C.C. Homes,$^3$ and A.J. Millis$^4$}
\affil{$^1$Laboratoire de Physique du Solide (UPR5 CNRS) ESPCI, 
75231 Paris cedex 05, France\\
$^2$Center for Superconductivity Research, Department of 
Physics, University of Maryland, College Park, Maryland 20742, 
USA.\\
$^3$ Department of Physics, Brookhaven National Laboratory, Upton, 
New York 11973, USA.\\
$^4$ Physics Department, Columbia University, New York, 
New York 10027, USA.
}

\begin{abstract}
We studied the optical conductivity of electron doped 
Pr$_{1-x}$Ce$_x$CuO$_4$ from the underdoped to the 
overdoped regime. The observation of low to high
frequency spectral weight transfer reveals the presence 
of a gap, except in the overdoped regime. A Drude peak 
at all temperatures shows the partial nature of this gap.
The close proximity of the doping at which the gap vanishes 
to the antiferromagnetic phase boundary leads us to assign
this partial gap to a spin density wave.

\end{abstract}


\section{Introduction}

The electron and hole doped cuprates phase diagram shows
a global symmetry. However, many aspects of the electron doped compounds, 
including the nature of the superconducting gap, the behavior of 
the normal state charge carriers, and the presence of a normal state 
(pseudo)gap are still unclear. A pseudogap phase is now well 
established on the hole doped side \cite{TimuskReview}. In 
Bi$_2$Sr$_2$CaCu$_2$O$_{8+\delta}$, angle resolved 
photoemission spectroscopy measurements (ARPES) indicate a pseudogap 
opening along the $(0,\pi)$ direction in $k$ space \cite{JCCetal}.
However, the in-plane optical conductivity does not show any direct 
evidence of this pseudogap \cite{Santander-Syro}. The optical 
conductivity 
of non superconducting Nd$_{2-x}$Ce$_x$CuO$_4$ (NCCO) 
single crystals ($x=0$ to $0.125$) suggests the opening of 
a high energy partial gap well above $T_{Neel}$ \cite{Onose}. 
Low temperature ARPES reveals a Fermi surface characterized by 
the presence of pockets \cite{Armitage}.

We determined the temperature evolution of the optical conductivity 
in a set of  Pr$_{2-x}$Ce$_x$CuO$_4$ thin films. 
Our data reveals the onset of a ``high energy'' partial gap below a 
characteristic temperature $T_W$ which evolves with doping. It is 
clearly detected for $0.13$, it is absent down to $20$ K for 
$x=0.17$ and it has a subtle signature for $x=0.15$ (optimal doping).
The proximity of our samples to the antiferromagnetic phase makes
a spin density wave (SDW) gap the natural interpretation for our 
observations, consistent with ARPES \cite{Armitage}.

\section{Experimental}

The thin films studied in this work were epitaxially grown by pulsed-laser
deposition on a SrTiO$_3$ substrate \cite{FournierSample}. The samples 
studied are (i) $x=0.13$ (underdoped) $T_c=15$ K (thickness 
3070~\AA), (ii) $x=0.15$ (optimally doped), $T_c=21$ K (thickness 3780~\AA)
and (iii) $x=0.17$ (overdoped) $T_c=15$ K (thickness 3750~\AA). All $T_c$'s 
were characterized by electrical resistance measurements. 
We checked the $x=0.15$ sample homogeneity by electron microscopy 
analysis (using the micron scale X-ray analysis of an EDAX system)
and found no dispersion at the micron scale in the Pr, Ce or Cu 
concentrations. Thin films are easy to anneal but, most important,
they can be made superconducting in the underdoped regime, whereas this 
seems difficult for crystals \cite{Onose}. Infrared-visible reflectivity 
spectra (at an incidence angle of $8^{\circ}$), were measured for all the 
films in the 25--21000~cm$^{-1}$ spectral range with a Bruker IFS-66v Fourier 
Transform spectrometer within an accuracy of 0.2\%. Typically 12 temperatures 
(controlled better than 0.2 K) were measured between 25 K and 300 K. 
The far-infrared frequency range (10--100~cm$^{-1}$) was measured for 
samples (ii) and (iii)  utilizing a Bruker IFS-113v at Brookhaven National 
Laboratory. 

\section{Results and Discussion} 

\begin{figure}
  \begin{center}
    \includegraphics[width=12cm]{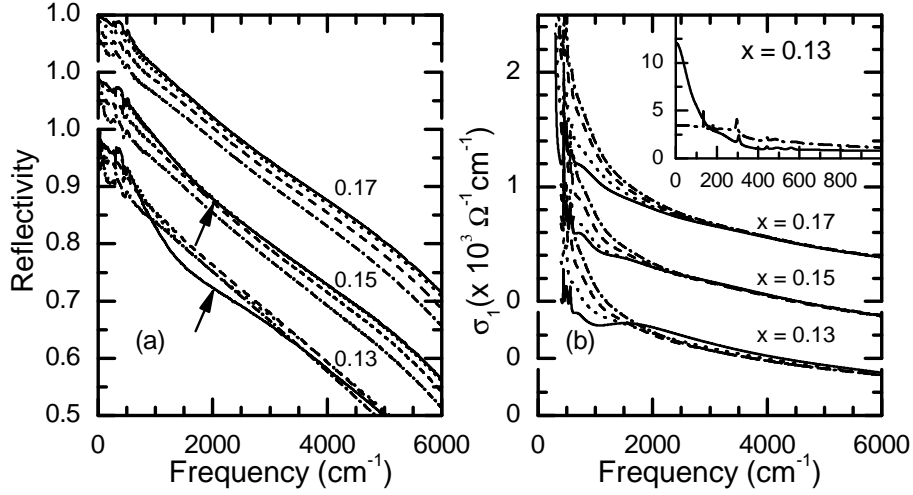} 
  \end{center}
\caption{(a) Infrared reflectivity of $x=0.13$, 0.15 and 0.17 
samples. Curves are shifted from one another by 0.1 for clarity.
(b) Real part of the optical conductivity from 400 to 6000 cm$^{-1}$. 
Curves are shifted by 400~$\Omega^{-1}$ cm$^{-1}$. The inset shows 
the low energy (0-1000~cm$^{-1}$) free carrier contribution to 
$\sigma_1(\omega)$ of the $x$=0.13 sample at 300~K and 25~K. In 
both panels the temperatures shown are 300~K (dash-dotted); 
200~K (dashed), 100~K (dotted) and 25~K(solid).}
\label{fig1}
\end{figure}

Figure \ref{fig1}(a) shows the raw reflectivity ($R$) from 25 to 
6000 cm$^{-1}$ for a set of selected temperatures. As the temperature 
decreases, an unconventional depletion of $R$ appears for 
$x=0.13$. This feature, denoted by an arrow, is still visible for $x=0.15$ 
as a subtle change in $R$. Conversely, the reflectivity of the $x=0.17$ sample 
increases monotonously with decreasing temperature over the whole spectral range 
shown. We applied a standard thin film fitting procedure to extract the 
optical conductivity from this data set \cite{Santander-Syro}. The real 
part $\sigma_1(\omega)$ of the optical conductivity is plotted in 
Fig. \ref{fig1} (b). At low energies, for all concentrations, the 
Drude-like contribution narrows as the temperature is lowered in the 
normal state from 300~K to 25~K (Fig. \ref{fig1} inset). This corresponds 
to a quasiparticle lifetime increasing in agreement to the metallic 
behavior of the resistivity. Figure \ref{fig1}(b) shows that the feature
in the reflectivity of the $x=0.13$ sample produces a dip/hump structure
in $\sigma_1$ with a peak at $\sim  1500$~cm$^{-1}$. For $x=0.15$ the 
reflectivity behavior is not clearly seen in $\sigma_1$.
A similar feature was observed in NCCO single crystals only for doping 
levels where such crystals are {\it not superconducting} \cite{Onose}, 
whereas we observe it in the $x$=0.13 sample. 

\section{Partial gap}

To understand the dip/hump structure, we define the partial sum rule
$W(\omega) = \int_0^\omega \sigma_1(\omega^\prime)d\omega^\prime$.
Making $\omega \rightarrow \infty$ yields the standard $f$-sum rule
$W =\pi n e^2 / 2 m$. When integrated over our full 
measured spectral range, we find a temperature independent
$W$ in all samples. 
Figure 2(a) shows the normalized temperature dependence 
$W(2000\textrm{ cm}^{-1},T)/W(2000\textrm{ cm}^{-1},300\textrm{ K})$ 
for all films. The continuous increase of $W$ with decreasing $T$ 
observed in the $x=0.17$ sample is a signature of decreasing scattering 
rate. In the $x=0.13$ sample $W(2000\textrm{ cm}^{-1})$ decreases for 
$T<150$~K, corresponding to the opening of a gap. As a Drude peak is 
present at all $T$'s, we conclude that the gap covers only part of the 
Fermi surface. The behavior of the $x=0.15$ sample is intermediate, 
suggesting a small or broadened gap.

\begin{figure}
  \begin{center}
    \includegraphics[width=12cm]{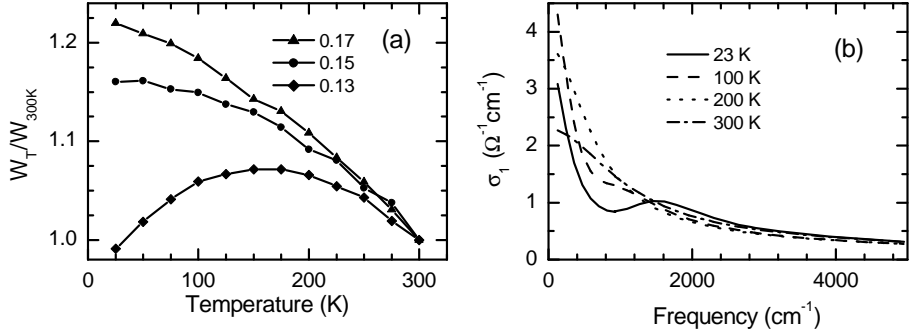} 
  \end{center}
\caption{(a) Temperature dependence of partial sum rule for samples with 
$x=0.13$, 0.15 and 0.17 integrating $\sigma_1$ up to 2000 cm$^{-1}$.
(b) Optical conductivity calculated by a spin density wave model
for $x=0.125$.}
\label{fig2}
\end{figure}

A possible interpretation for the origin of the gap is a commensurate 
$(\pi,\pi)$ spin density wave. It induces a symmetry breaking, 
folding the Fermi surface upon itself, and a partial gap $\Delta_{SDW}$ 
opens at the intersection of the antiferromagnetic Brillouin zone, creating 
pockets in the Fermi surface \cite{Armitage}.

Figure \ref{fig2}(b) shows calculations \cite{Millis} using a Marginal 
Fermi liquid with parameters chosen to reproduce $\rho(T)$ for $T>200$~K, 
combined with a commensurate $(\pi,\pi)$ SDW gap opening for $T<200$~K. 
The $T=0$ gap magnitude was adjusted to correctly locate the maximum 
in $\sigma$ at $T=0$. The calculation is seen to reproduce the data 
fairly well [compare to Fig. \ref{fig2}(b)].

\section{Summary}

We have measured with great accuracy the reflectivity of 
electron doped Pr$_{2-x}$Ce$_x$CuO$_4$ at various Ce doping 
levels. An optical conductivity spectral weight analysis shows that
a partial gap opens at low temperatures for Ce concentrations up to 
$x=0.15$. A spin density wave model reproduces satisfactorily the data.

{\footnotesize
\vskip0.2cm
\noindent Research supported in part by NSF-DMR-0102350 (Maryland) and
NSF-DMR-0338376 (Columbia).
}

\end{document}